\documentclass[letterpaper]{article}
\usepackage{aaai2026}
\usepackage{times}
\usepackage{helvet}
\usepackage{courier}
\usepackage[hyphens]{url}
\usepackage{amsmath}
\usepackage{graphicx}
\usepackage{natbib}
\usepackage{caption}
\usepackage{newfloat}
\usepackage{listings}
\usepackage{booktabs}

\DeclareCaptionStyle{ruled}{labelfont=normalfont,labelsep=colon,strut=off}
\nocopyright
\title{Mapping Election Toxicity on Social Media across\\Issue, Ideology, and Psychosocial Dimensions}

\author{
    Lei Cao,
    Wen Zeng,
    Xinyue Wu,
    Eun Cheol Choi, 
    Emilio Ferrara 
}
\affiliations{
    University of Southern California\\
    
    \{caolei, zengwen, xwu11613, euncheol, emiliofe\}@usc.edu
}

\usepackage{xcolor}

\definecolor{myyellow}{RGB}{204,153,0}
\definecolor{myblue}{RGB}{66, 133, 244}

\begin{document}

\maketitle

\begin{abstract}

Online political hostility is pervasive, yet it remains unclear how toxicity varies across campaign issues and political ideology, and what psychosocial signals and framing accompany toxic expression online. In this work, we present a large-scale analysis of discourse on X (Twitter) during the five weeks surrounding the 2024 U.S. presidential election. We categorize posts into 10 major campaign issues, estimate the ideology of posts using a human-in-the-loop LLM-assisted annotation process, detect harmful content with an LLM-based toxicity detection model, and then examine the psychological drivers of toxic content. We use these annotated data to examine how harmful content varies across campaign issues and ideologies, as well as how emotional tone and moral framing shape toxicity in election discussions. Our results show issue heterogeneity in both the prevalence and intensity of toxicity. Identity-related issues displayed the highest toxicity intensity. As for specific harm categories, \textit{harassment} was most prevalent and intense across most of the issues, while \textit{hate} concentrated in identity-centered debates. Partisan posts contained more harmful content than neutral posts, and ideological asymmetries in toxicity varied by issue. In terms of psycholinguistic dimensions, we found that toxic discourse is dominated by high-arousal negative emotions. Left- and right-leaning posts often exhibit similar emotional profiles within the same issue domain, suggesting emotional mirroring. Partisan groups frequently rely on overlapping moral foundations, while issue context strongly shapes which moral foundations become most salient. These findings provide a fine-grained account of toxic political discourse on social media and highlight that online political toxicity is highly context-dependent, underscoring the need for issue-sensitive approaches to measuring and mitigating it.

\end{abstract}

\section{Introduction}

The 2024 U.S. presidential election has unfolded in an increasingly polarized digital environment, where social media platforms such as X (formerly Twitter) serve as central arenas for political debate \citep{samsir2024machine,mittal2025evolution}. While these platforms enable large-scale participation, they also facilitate the spread of hostile and toxic discourse, including harassment, hate speech, and antagonistic interactions \citep{efstratiou_2023_nonpolar,garland_2020_countering,rossini_2020_beyond}. Understanding how toxicity evolves in real-time election discourse is, therefore, critical for assessing the quality of online political communication.

Prior research shows that online political discourse is deeply shaped by polarization and platform dynamics \citep{abramowitz_2008_is,barbera2015birds}. Algorithmic curation and selective exposure reinforce echo chambers and affective polarization, fostering antagonistic interactions between ideological groups \citep{efstratiou_2023_nonpolar,iyengar2019origins}. In addition, political hostility varies across topics. Research on election discourse highlights issue salience and distinguishes between \textit{valence issues}, which generate broad consensus, and \textit{position issues}, which activate identity and moral conflict and drive polarization \citep{alt_1979_the,stokes_1963_spatial}. Consistent with this, computational studies of harmful speech have documented the large-scale prevalence of hate and incivility in political discussions, showing that toxicity varies systematically across topics and often intensifies around salient events and controversial debates \citep{salminen_2020_topicdriven,rao_2025_polarized}. Beyond structural and topical factors, psychological processes further shape these dynamics: high-arousal negative emotions spread rapidly in online networks \citep{brady2017emotion}, while moralized issue framing transforms disagreement into conflicts between competing value systems \citep{clifford2015moral}. Together, these perspectives suggest that toxicity emerges from the interaction of issue context, ideology, and psychosocial mechanisms.

However, existing studies often treat toxicity as a general phenomenon, overlooking how it varies across specific political issues and ideological groups \citep{garland_2020_countering,rao_2025_polarized}. Additionally, although prior work has explored ideology, emotional expression, and moral foundations across different topics \citep{dechoudhury2016social,gallagher2018divergent}, these dimensions are typically studied in isolation and rarely integrated within an issue-specific framework or systematically compared across multiple issues. To address these gaps, we proposed the following research questions.

\begin{itemize}
    \item \textbf{RQ1} (Issue Context): How do the prevalence and severity of toxicity vary across different major campaign issues during the 2024 U.S. election?
    
    \item \textbf{RQ2} (Ideology): Within each campaign issue, how does toxic content differ between political ideologies?
    
    \item \textbf{RQ3} (Psychological Mechanism): What emotional expressions and moral foundations characterize toxic discourse across different campaign issues and ideologies?
\end{itemize}

To answer these research questions, this study analyzes 726,566 original posts from X during the final stage of the 2024 U.S. election. Using a human-in-the-loop, LLM-assisted annotation framework, we labeled each post with the campaign issues and political ideology mentioned. We also detected toxicity using an LLM-based detection model and used existing BERT-based models to detect emotion and moral foundations. We integrated all of these dimensions of analysis to create a high-resolution map of online toxicity during the election.

Our findings show that toxicity is highly issue-dependent, with identity-related topics exhibiting higher prevalence and intensity. We also find that partisan discourse is consistently more toxic than neutral content, with ideological differences varying across issues. Finally, toxic discourse is dominated by high-arousal negative emotions and issue-specific moral framing, indicating that underlying psychosocial mechanisms structure hostility. By linking issue context, ideology, and psychological mechanisms, this study advances a more integrative, high-resolution framework for understanding online political hostility.

\section{Related Work}

\subsection{Toxicity in Political Discourse on Social Media}

Political discourse on social media is increasingly characterized by toxicity. Political polarization is typically defined as the extent of ideological division between groups across the political spectrum \cite{dimaggio_1996_have,abramowitz_2008_is}, and can be understood both as a state and as a process of intensification over time.

On social media, patterns of information exposure are often uneven, with engagement concentrated among a subset of users interacting with partisan or low-quality content \cite{guess_2020_exposure}. Platform affordances further shape how political interaction unfolds, often privileging antagonistic over deliberative exchange \cite{an_2024_curated}. Algorithmic curation and selective exposure reinforce these dynamics by placing users in ideologically homogeneous networks, commonly described as echo chambers \cite{barbera2015birds,garimella2018political}. Such environments contribute to affective polarization, characterized by increased dislike and distrust toward political outgroups \cite{iyengar2019origins}, a pattern also reflected in higher levels of toxicity in cross-group interactions \cite{lerman_2024_affective}.

Within these conditions, toxicity emerges as a systematic feature of political communication rather than an isolated phenomenon. Toxic language functions not only as an expression of disagreement but also as a marker of group identity, reinforcing in-group cohesion while targeting out-groups \cite{waseem2016hateful,davidson2017automated}. These dynamics are further amplified by networked interaction, where engagement within aligned communities is associated with increased antagonism toward others \cite{efstratiou_2023_nonpolar}, and by discourse patterns that frame outgroups as threatening or morally inferior \cite{gerard_2025_fear}. As a result, political disagreement is more likely to manifest in toxic rather than deliberative forms.

Despite extensive research on polarization, platform-mediated exposure, and online toxicity, existing work has largely examined these dynamics in isolation and focused on the overall prevalence of toxic discourse. As a result, it remains unclear how toxicity varies across political issues and why certain topics are more likely to generate toxic expression than others. This study addresses this gap by examining issue-level variation in toxicity and linking patterns of toxic expression to differences in political topics within social media discourse.

\subsection{Divisive Issues in U.S. Election}

Political hostility on social media is not evenly distributed across topics. Research emphasizes the role of issue salience, the perceived importance of an issue, in shaping patterns of engagement and conflict \cite{berelson_1986_voting,wilhelm_2024_a}. Scholars distinguish between \textit{valence issues}, which evoke broad consensus (e.g., economic growth) \cite{stokes_1963_spatial}, and \textit{position issues}, which activate deeply rooted social identities and moral values, such as abortion, gun control, and immigration \cite{alt_1979_the}. These issue differences are closely related to issue ownership, whereby political actors are perceived to “own” specific issues and maintain stable positions on them \cite{budge_2025_explaining, petrocik_1996_issue}. When competing groups contest ownership over such issues, political disagreement becomes more intense and polarized \cite{dedreu_2005_the, wilhelm_2024_a}. These position issues are particularly potent drivers of polarization, as they frame political disagreement as a clash between incompatible moral worldviews rather than competing policy preferences \cite{mason2018uncivil, iyengar2019origins}.

Recent studies have also documented the large-scale prevalence of hate speech in online political discussions \cite{garland_2020_countering}. Computational analyses demonstrate that hostile and uncivil discourse is not uniformly distributed, but varies systematically across topics \cite{salminen_2020_topicdriven, kim2021incivility}. Studies of large-scale online discussions find that certain topics, particularly those involving sensitive social or political themes, are associated with higher levels of hostile and uncivil discourse, while others remain comparatively less contentious \cite{salminen_2020_topicdriven}. Moreover, prior work finds that incivility often spikes around controversial policy debates and politically salient events, highlighting the dynamic and context-dependent nature of toxic discourse \cite{theocharis_2020_the}. Complementing this, \cite{rossini_2020_beyond} shows that different discussion contexts are associated with distinct forms of incivility and intolerance, further underscoring the role of contextual factors in shaping online political conflict. Furthermore, event-centered studies illustrate this pattern, showing that discourse surrounding highly salient and contentious issues, such as racial justice movements or gun control debates, exhibits markedly higher levels of hostility and emotional intensity \cite{dechoudhury2016social, gallagher2018divergent, rao_2025_polarized, zaman_2025_disagreement}.

Despite these advances, several limitations remain. Although prior work demonstrates that toxicity varies across topics, most studies operationalize topics using broad, domain-general taxonomies, rather than election-specific issue categories \cite{theocharis_2020_the, salminen_2020_topicdriven}. This may limit their ability to capture how toxic discourse is organized around campaign-relevant issues that structure real-world electoral conflict. Besides, while computational approaches have enabled large-scale measurement of harmful speech, they often focus on aggregate patterns \cite{garland_2020_countering}, overlooking how toxicity interacts with political ideology within issue domains. These limitations highlight the need for an issue-centered, multidimensional analysis that anchors toxicity to election-specific campaign issues while jointly examining its interactions with ideology and fine-grained measures of harmful expression. 

\subsection{Psychological Drivers of Online Toxicity}

Beyond structural and topical factors, psychological processes play a central role in shaping how toxicity is expressed in online political discourse. The rapid and fragmented nature of social media privileges intuitive, emotionally driven responses, increasing the likelihood of affective escalation. Emotion functions as a primary driver of toxicity, as judgments are often guided by intuitive rather than analytic processes \cite{haidt_2001_the}. These dynamics are amplified online, where emotional content spreads quickly and receives disproportionate engagement \cite{kramer2014experimental,brady2017emotion}, often producing cycles of escalating negative affect across opposing groups \cite{rao2023affective}.

These emotional responses are structured through processes of social identity. Individuals align their evaluations with group norms and interpret disagreement as intergroup conflict \cite{turner_1987_rediscovering,abrams_2004_metatheory}. Under such conditions, toxic expression serves both reactive and performative functions, responding to opposing views while signaling alignment with one’s political group. Social media further intensifies these dynamics through visibility, networked interaction, and ideological clustering, leading cross-group exchanges to manifest in more toxic forms \cite{an_2024_curated,engel_2025_social} 

Building on this, Moral Foundations Theory explains why toxicity varies across political issues. Moral judgments are grounded in intuitive value systems that differ across groups \cite{graham2009moral,graham2013moral}, and political disagreement often reflects conflicts between these moral frameworks rather than factual disputes \cite{clifford2015moral}. In online discourse, moral values are mobilized as boundary-making resources that construct simplified “us versus them” distinctions, amplifying emotional engagement and perceived moral certainty \cite{wickenkamp_2025_the}. These dynamics are observable at scale, where groups rely on distinct moral framings that intensify debate \cite{mejova_2023_authority}. As a result, when issues activate deeply held moral values, disagreement becomes moralized, reducing the likelihood of compromise and increasing the perceived legitimacy of toxic expression.

Taken together, toxicity in online political discourse is not incidental but systematically produced through the interaction of emotional intensity, social identity, and moralization under social media conditions. However, less is known about how psychological factors such as emotion and moral framing shape these dynamics at scale across different issue contexts. This study examines issue-level variation in toxicity and integrates ideological, emotional, and moral dimensions in the analysis of large-scale election discourse on social media.

\section{Methods}
\subsection{Data}
We use a large-scale X dataset from \citep{balasubramanian2024publicdatasettrackingsocial}, covering May 1, 2024, to November 30, 2024. Relying on targeted keywords linked to key political figures, events, and emerging issues in U.S. election discourse, this dataset comprises 46 million publicly available posts on X about the run-up to the 2024 U.S. Presidential Election. From this dataset, we exclude retweets, replies, and quoted tweets to focus on original tweets, i.e., content generated exclusively by users. Also, the data range from Oct 21, 2024, to Nov 24, 2024, covering the five weeks of the heated final phase, Election Day on Nov 5, 2024, and the subsequent weeks of reactions to the election results. This leaves us with 726,566 original tweets during the five weeks. 

\subsection{Toxicity}

We detected harmful content in the dataset using OpenAI's \texttt{omni-moderation-latest} endpoint \cite{kivlichan2024upgrading}. This model provides a binary overall toxic label and identifies 14 distinct harm categories, including six main categories and eight subcategories. Each category is associated with a continuous score ranging from 0 to 1, representing the model’s confidence that the text contains such content.
A post is flagged as toxic if any harm category is flagged as true. In this study, we focus on six main toxic categories, \textit{harassment}, \textit{hate}, \textit{violence}, \textit{illicit},  \textit{self-harm}, \textit{sexual}, as the subcategories are too detailed for this study.

\subsection{Campaign Issue and Political Ideology Annotation}

\subsubsection{LLM-Assisted Annotation} Following recent work on LLM-assisted annotation in computational social science \cite{tornberg2024best,tornberg2025llm,pangakis2024humancentered}, we adopted a human-in-the-loop workflow to annotate both campaign issues and political ideology in election-related posts. Specifically, we combined human annotation, iterative codebook refinement, and LLM-based scaling. We began by developing an initial codebook and prompt for: (1) identifying the primary campaign issue discussed in each post and (2) identifying the ideological signal expressed in the post text. To construct the initial issue schema, three annotators first reviewed resources on major campaign issues in the 2024 U.S. election and collaboratively defined a preliminary codebook. Next, three human annotators independently annotated a random sample of 100 posts. During this stage, we also recorded model-generated labels to facilitate comparison and prompt refinement. The annotators then met to review disagreements, clarify ambiguous cases, and revise the codebook and prompts accordingly. This process was especially important for issue categories with broad meanings, such as \textit{Democracy} and \textit{Economy}, and for ideology labels where posts contained weak or implicit signals. The labels with majority agreement became the ground truth for further LLM annotation. To scale annotation to the full corpus, we used Qwen3.5-9B\footnote{https://huggingface.co/Qwen/Qwen3.5-9B} for issue classification and Llama 3.2-3B\footnote{https://huggingface.co/meta-llama/Llama-3.2-3B} for ideology detection, with structured prompts developed through an iterative coding process. We selected these models for their balance of performance, inference efficiency, and cost in large-scale annotation settings. For ideology detection, we used a US-developed model because prior work suggests that LLM outputs vary with the political and cultural context of model development \cite{buyl2026large}. We deployed the model on a NVIDIA L40S GPU and used vLLM \cite{kwon2023efficient} with asynchronous inference to accelerate large-scale annotation.

\subsubsection{Campaign Issue} Although issue classification has a long tradition in policy research, including the Policy Agendas Project codebook \cite{jones2023pap}, campaign issues follow a more strategic and time-sensitive logic. In electoral campaigns, parties and candidates selectively emphasize issues in response to voter priorities, competitive pressures, and unfolding events, rather than reproducing stable institutional policy domains \cite{kluver2016agenda,meyer2016issue,druckman2009campaign}. Issue emphasis can also shift across campaigns as parties adapt to electoral opportunity structures and issue ownership dynamics \cite{greenspeding2019party,walgrave2014limits,neundorf2018microfoundations}. As a result, campaign discourse is more fluid and context-dependent than formal policy discourse. Thus, we do not rely on existing policy topic taxonomies to annotate campaign issues. 

We operationalized the annotation as single-label classification by assigning each post its most relevant issue. Referring to Policy Agendas Project codebook \cite{jones2023pap} and Pew Research Center's typology of major U.S. political issues during 2024 election\footnote{\url{https://www.pewresearch.org/politics/2024/09/09/issues-and-the-2024-election/}}, we identified ten campaign issue categories: \textit{Economy}, \textit{Healthcare \& Welfare}, \textit{Immigration}, \textit{Abortion}, \textit{Climate Change}, \textit{Violent Crime}, \textit{Gun Policy}, \textit{Foreign Policy}, \textit{Race \& Gender Inequality}, and \textit{Democracy} (see Appendix~A). This scheme was intended to maximize substantive interpretability for the 2024 election context rather than reproduce a universal policy taxonomy. 

\subsubsection{Political Ideology}
Based on the ideological signal that could be inferred from the post text, each post received one of three labels: \textit{left}, \textit{right}, or \textit{neutral}. This task focuses on ideology as expressed in the text, rather than on user-level partisanship. Posts were labeled \textit{neutral} when they did not convey a clear ideological leaning, including posts that were primarily informational, ambiguous, or lacking sufficient textual cues. 

\subsection{Psychological Dimensions of Toxic Content}

We filter the dataset to include only posts identified as toxic content for a deeper exploration of the psychological drivers behind toxicity in the online political discussion. 

\subsubsection{Emotion}

We used the fine-tuned RoBERTa-based model \texttt{emotion-english-distilroberta-base}\footnote{\url{https://huggingface.co/j-hartmann/emotion-english-distilroberta-base}} to detect emotional content. According to the model card, this checkpoint was trained on six diverse English emotion datasets: the CrowdFlower Emotion in Text dataset \cite{crowdflower2016emotion} and the Twitter emotion dataset, \cite{saravia2018carer}, GoEmotions \cite{demszky2020goemotions}, ISEAR \cite{scherer1994universality}, MELD \cite{poria2019meld}, and SemEval-2018 Task 1: Affect in Tweets \cite{mohammad2018semeval}. The model predicts the six basic Ekman\textit{ emotions: joy}, \textit{sadness}, \textit{disgust}, \textit{fear}, \textit{surprise}, and \textit{anger}, plus a \textit{neutral} class \cite{ekman1992argument,hartmann2022emotion}. Given an input text, the model returns a score between 0 and 1 for each of the seven labels. 

\subsubsection{Moral Foundation}

We measured moral language using MoralBERT, a set of transformer-based models fine-tuned to detect moral rhetoric in social media text \cite{preniqi2024moralbert}. 
Moral Foundations Theory (MFT) posits that moral reasoning is structured around a set of core foundations: \textit{Care}/\textit{Harm}, \textit{Fairness}/\textit{Cheating}, \textit{Loyalty}/\textit{Betrayal}, \textit{Authority}/\textit{Subversion}, and \textit{Sanctity}/\textit{Degradation} \cite{graham2009moral}.
Following MFT, MoralBERT was trained on multiple social media datasets with human annotations, including the Moral Foundations Twitter Corpus, the Moral Foundations Reddit Corpus, and Facebook vaccination posts \cite{preniqi2024moralbert,hoover2020mftc,trager2022mfrc}. Compared with lexicon-based approaches, such as the Moral Foundations Dictionary (MFD) \cite{graham2009moral}, MoralBERT has been shown to better capture context-dependent moral meaning in social discourse \cite{preniqi2024moralbert}. In our study, we used the set of models to identify the five pairs of moral foundations. For each post, the models returned ten labels with corresponding scores between 0 and 1 representing the model's confidence that the text contains content related to the moral foundations.

\section{Results}
Before addressing our research questions, we provide a descriptive overview of the labeled dataset to understand the distribution of toxicity, campaign issues, and political ideologies during the observed period. 

\subsubsection{Toxicity}

Out of the total corpus, our pipeline flagged 110,413 posts as containing toxic content, representing 15.2\% of the dataset. Table \ref{tab:descriptive_analysis_toxicity} shows the distribution of toxicity. \textit{Harassment} (11.8\%) and \textit{Violence} (4.5\%) were the most prevalent categories. In terms of toxicity intensity, \textit{Harassment} had the highest average toxicity score ($M = 0.69$, $SD = 0.18$). Although less frequent (1.8\%), \textit{Hate} also showed relatively high scores ($M = 0.62$, $SD = 0.14$). Additional descriptive statistics for all toxicity categories are reported in Appendix~B.

\subsubsection{Campaign Issues}

The inter-rater reliability, measured by Krippendorff's $\alpha$ \cite{krippendorff2004reliability}, among the three annotators was 0.84, indicating strong agreement in issue annotation. On the human-annotated validation set, the LLM issue classifier achieved a weighted F1 score of 0.81. LLM-assisted issue annotation identified 435,316 posts assigned to campaign issues, representing 59.9\% of the corpus. \textit{Democracy} was the most frequent issue (28.8\%), followed by \textit{Foreign Policy} (14.1\%). \textit{Economy} (5.3\%) and \textit{Immigration} (4.3\%) were also prominent, while the remaining issues were less frequent, including \textit{Racial \& Gender Inequality} (2.8\%), \textit{Healthcare \& Welfare} (2.4\%), \textit{Climate Change} (0.8\%), \textit{Abortion} (0.7\%), \textit{Violent Crime} (0.4\%), and \textit{Gun Policy} (0.2\%).

\subsubsection{Political Ideology}

In terms of political ideology annotation, the LLM ideology classifier achieved an F1 score of 0.80 on the human-annotated validation set. Based on ideological signals inferred from post text alone, we classified posts into three categories: \textit{left}, \textit{right}, and \textit{neutral}. The distribution reveals a clear asymmetry in the corpus. Right-leaning discourse was substantially more prevalent in the dataset during the observation period (49.7\%), followed by \textit{neutral} posts (33.1\%), and \textit{left}-leaning posts (17.1\%). 

\subsection{RQ1: Toxicity Across Campaign Issues}

To answer our RQ1 regarding the distribution of toxicity across campaign issues, we analyzed its prevalence across 10 distinct campaign issues. The results reveal significant heterogeneity in how toxicity manifests across issues under debate.

\subsubsection{Toxicity Prevalence}

We calculated the percentage of posts flagged as toxic within each issue (see Appendix~B). Issues involving physical safety, rights, and identity tended to exhibit relatively high levels of toxicity. \textit{Violent Crime} emerged as the most toxic issue, with 57.8\% of posts containing toxic content. It was followed by \textit{Racial \& Gender Inequality} (39.2\%), \textit{Immigration} (29.2\%), and \textit{Abortion} (24.2\%). \textit{Gun Policy} (21.2\%) and \textit{Foreign Policy} (21.0\%) also showed elevated toxicity prevalence. In contrast, \textit{Climate Change} exhibited the lowest toxicity rate (5.2\%), followed by \textit{Economy} (5.6\%), \textit{Healthcare \& Welfare} (8.0\%).

\begin{figure}[htbp]
    \centering
    \includegraphics[width=1\linewidth]{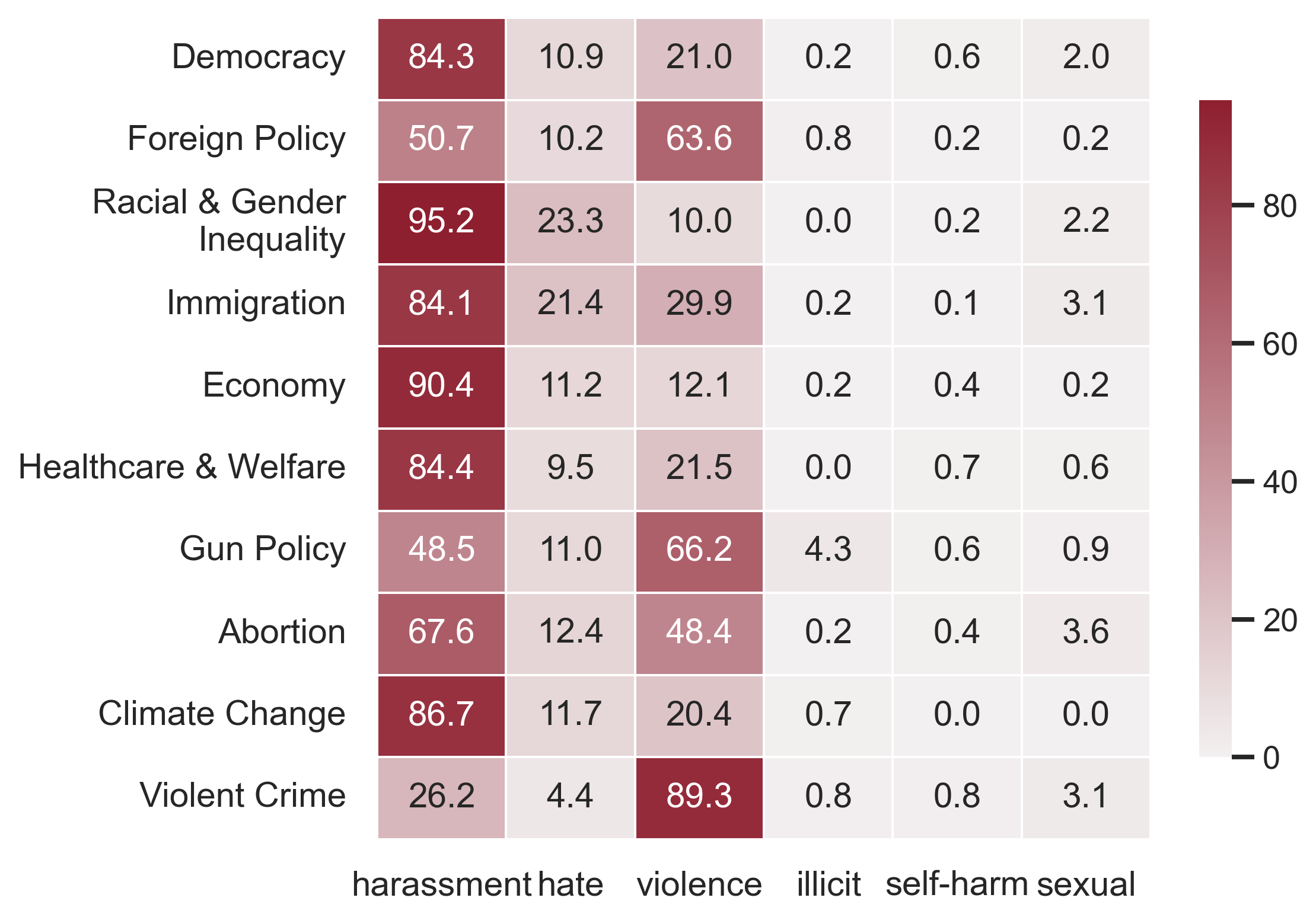}
    \caption{Toxicity Category Composition by Issue}
    \label{fig:rq1_issue_toxicity_category_heatmap}
\end{figure}

Beyond overall toxicity prevalence, the composition of toxicity also varies across issues. Figure \ref{fig:rq1_issue_toxicity_category_heatmap} breaks toxic posts down into specific toxicity categories.\footnote{Toxicity detection was conducted as a multi-label classification task, meaning that a single post could be assigned to more than one toxicity category. As a result, the category shares within an issue can sum to more than 100\%.} Three broad patterns emerge. 
First, \textit{Harassment} dominates most discussions of issues. It is the largest toxicity category in posts about \textit{Democracy}, \textit{Racial \& Gender Inequality}, \textit{Immigration}, \textit{Economy}, \textit{Healthcare \& Welfare}, \textit{Abortion}, and \textit{Climate Change}. 
Second, \textit{Violence} is especially prominent in security-related issues, reaching its highest levels in \textit{Violent Crime}, \textit{Foreign Policy}, and \textit{Gun Policy}. 
Third, although \textit{Hate} is less prevalent overall, it is comparatively more prominent in identity-related issues. In particular, \textit{Hate} reaches its highest shares in \textit{Racial \& Gender Inequality} and \textit{Immigration}. By contrast, the remaining three categories, \textit{Illicit}, \textit{Self-harm}, and \textit{Sexual}, account for small portions across issues, with less than 5\% each.

\subsubsection{Toxicity Intensity}

\begin{figure}[htbp]
    \centering
    \includegraphics[width=1\linewidth]{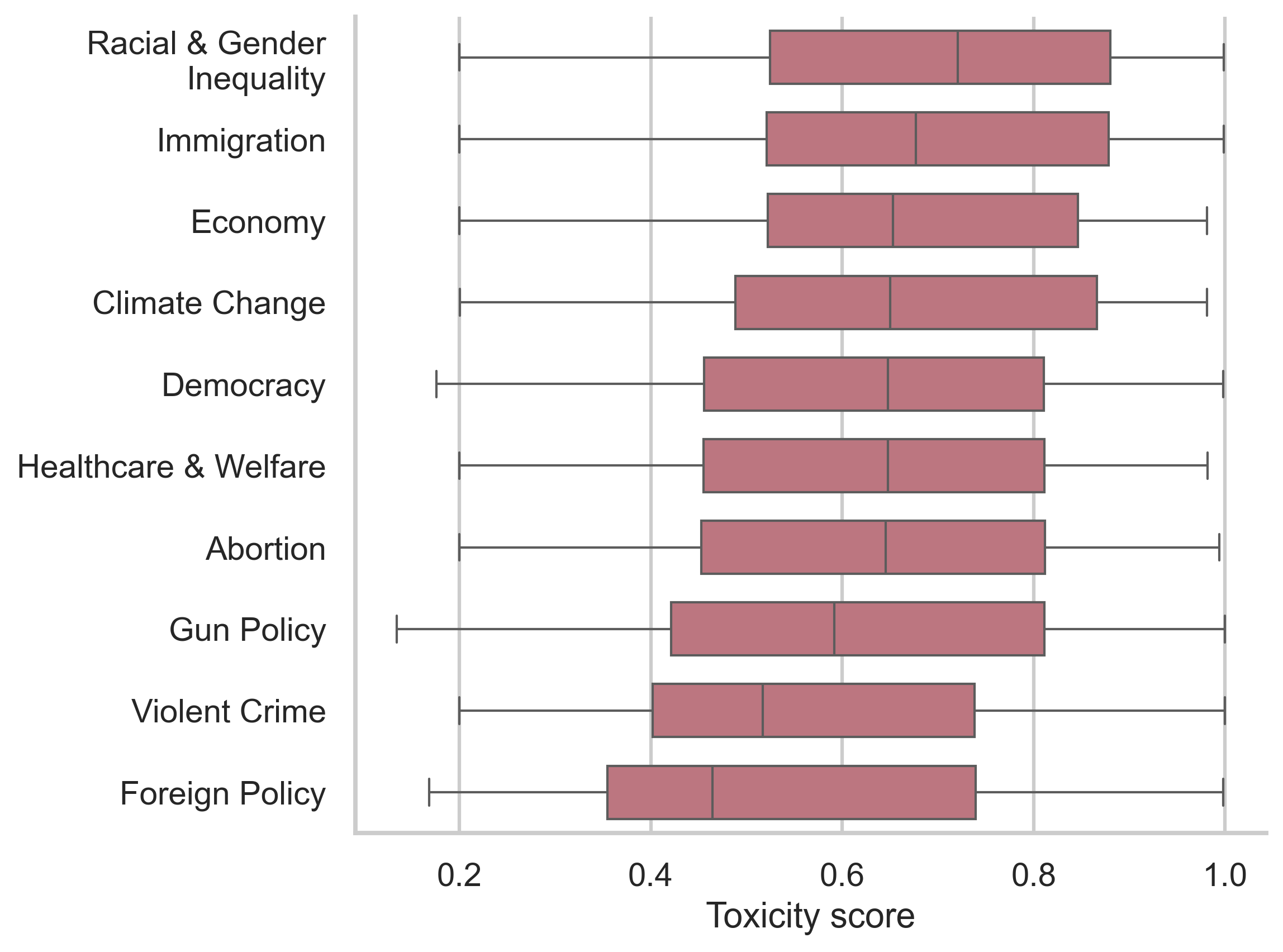}
    \caption{Toxicity Intensity by Issue}
    \label{fig:rq1_issue_toxicity_intensity_boxplot}
\end{figure}

Whereas toxicity prevalence captures how often toxic content appears within an issue, the toxicity score reflects the intensity of that content. Figure~\ref{fig:rq1_issue_toxicity_intensity_boxplot} shows the distribution of toxicity scores for toxic posts across the ten campaign issues, revealing both alignment and divergence between toxicity frequency and severity. First, identity-related issues exhibit the highest toxicity intensity. \textit{Racial \& Gender Inequality} ($M = 0.69$, $SD = 0.20$) and \textit{Immigration} ($M = 0.68$, $SD = 0.21$) showed the highest toxicity scores. This finding suggests that toxic discourse surrounding identity and migration is not only frequent but also intense. Second, several issues associated with violence or public safety, including \textit{Foreign Policy} ($M = 0.53$, $SD = 0.24$), \textit{Gun Policy} ($M = 0.60$, $SD = 0.22$), and \textit{Violent Crime} ($M = 0.55$, $SD = 0.21$), showed comparatively lower toxicity scores despite containing substantial amounts of toxic content. These patterns suggest that toxic content in these domains may more often be tied to descriptions of war, crime, or weapons. By contrast, toxicity in identity-centered discussions is more likely to take the form of direct hostile expression.

\subsection{RQ2: Ideology and Toxicity Within Issues}

To examine how toxicity varies by political orientation, we compared both the prevalence and the intensity of toxic content across posts labeled \textit{left}, \textit{right}, and \textit{neutral}. We also used Kruskal--Wallis tests to examine whether toxicity scores differed significantly across ideological groups within each toxicity category. Our findings suggest that ideological polarization in toxicity is issue-dependent rather than uniform. \textit{Neutral} posts are consistently the least toxic, while partisan discourse is more likely to contain toxic content. However, the direction and magnitude of ideological differences vary across issues. Some domains exhibit elevated toxicity on both sides, whereas others are marked by stronger toxicity prevalence or intensity among one ideological group.

\begin{figure}[htbp]
    \centering
    \includegraphics[width=1\linewidth]{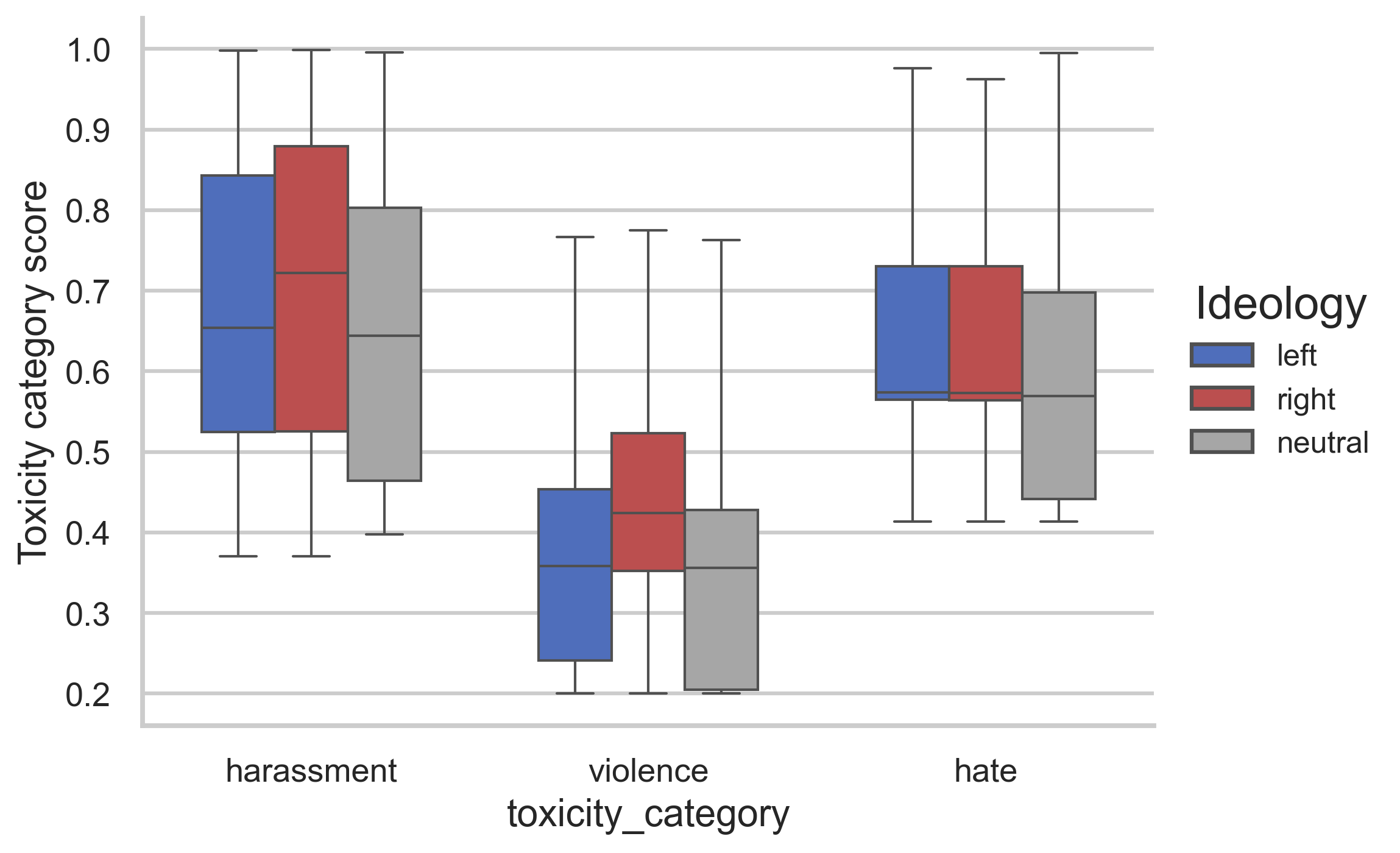}
    \caption{Toxic Category Scores by Ideology}
    \label{fig:rq2_ideology_top_category_scores_boxplot}
\end{figure}

In general, partisan posts exhibit a higher prevalence of toxicity (\textit{left}: 18.9\%; \textit{right}: 21.4\%) than neutral posts (4.0\%). Similar to toxicity intensity, Figure \ref{fig:rq2_ideology_toxicity_intensity_boxplot} shows that \textit{neutral} posts have the lowest toxicity scores overall ($M = 0.50$, $SD = 0.21$), whereas both \textit{left}-leaning ($M = 0.61$, $SD = 0.22$) and \textit{right}-leaning ($M = 0.66$, $SD = 0.21$) posts show significantly higher toxicity intensity ($p < 0.05$). This pattern suggests that toxicity is more strongly associated with ideologically inflected political expression.

At the category level, our analysis focused on the three most prevalent toxicity categories: \textit{harassment}, \textit{hate}, and \textit{violence}. Figure \ref{fig:rq2_ideology_top_category_scores_boxplot} shows that \textit{harassment}, the most prevalent toxicity category, also has the highest toxicity scores across all three ideological groups. Specifically, \textit{harassment} scores are highest among \textit{right}-leaning posts ($M = 0.70$, $SD = 0.19$), followed by \textit{left}-leaning posts ($M = 0.69$, $SD = 0.18$), and lowest among \textit{neutral} posts ($M = 0.62$, $SD = 0.17$). \textit{Hate} also exhibits relatively high scores across groups, although ideological differences are more modest that \textit{right}-leaning ($M = 0.62$, $SD = 0.14$) and \textit{left}-leaning ($M = 0.62$, $SD = 0.14$) posts both score slightly higher than \textit{neutral} posts ($M = 0.59$, $SD = 0.13$). By contrast, \textit{violence} has lower scores overall, but the ideological gap is more pronounced. \textit{Right}-leaning posts show the highest \textit{violence} scores ($M = 0.46$, $SD = 0.19$), compared with \textit{left}-leaning ($M = 0.40$, $SD = 0.18$) and \textit{neutral} posts ($M = 0.37$, $SD = 0.15$). Taken together, these results suggest that ideological differences are especially pronounced for \textit{harassment} and \textit{violence}, while \textit{hate} displays comparatively similar intensity across partisan groups but remains lower among \textit{neutral} posts.

\begin{figure}[htbp]
    \centering
    \includegraphics[width=1\linewidth]{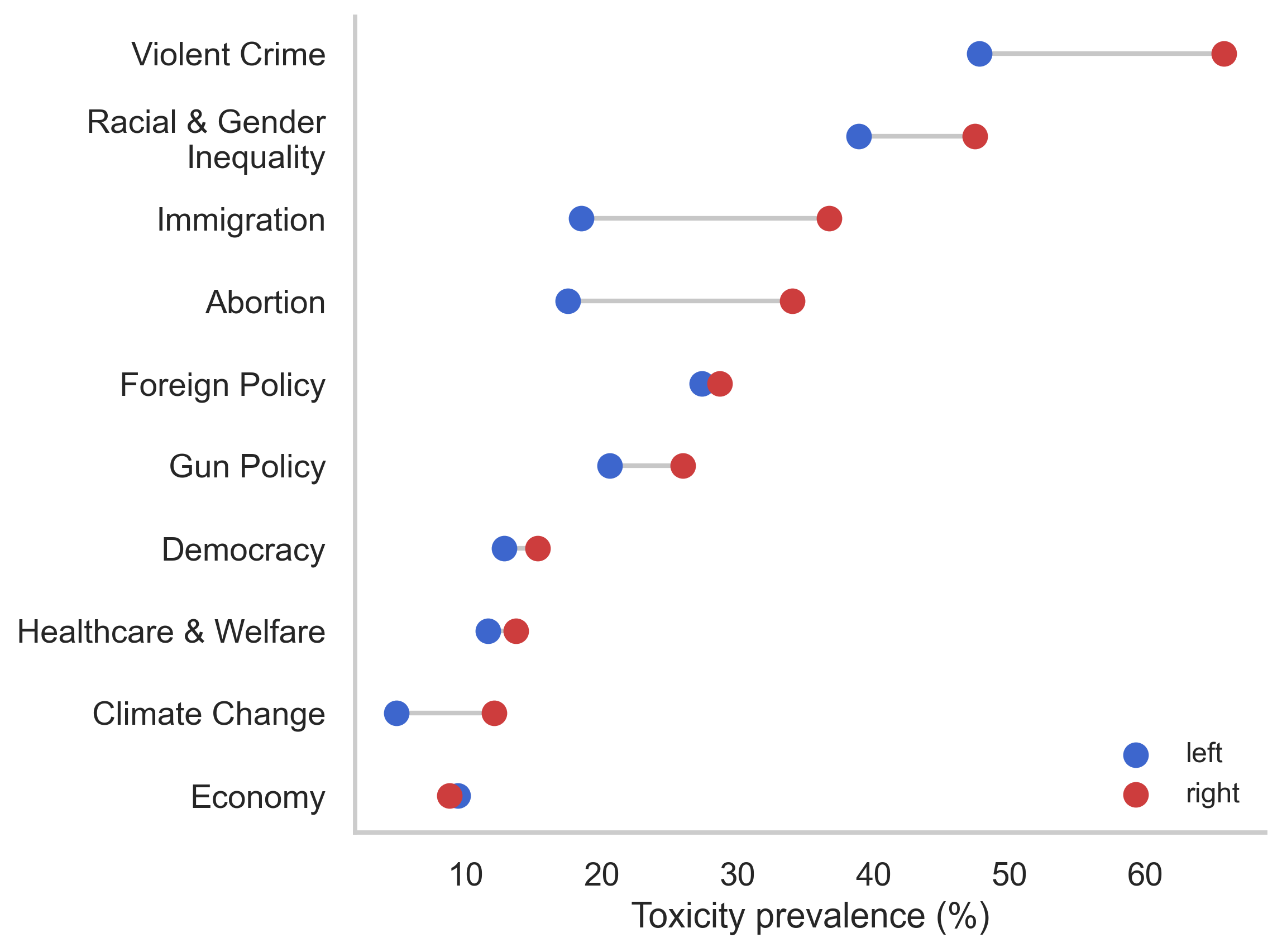}
    \caption{Toxicity Prevalence by Issue and Ideology}
    \label{fig:rq2_issue_ideology_toxicity_prevalence_dumbbell}
\end{figure}

We next examined whether ideological differences in toxicity vary across campaign issues. Figure~\ref{fig:rq2_issue_ideology_toxicity_prevalence_dumbbell} shows issue-level toxicity prevalence for \textit{left}-, \textit{right}-, and \textit{neutral} posts. Overall, \textit{right}-leaning posts exhibited higher toxicity prevalence in most issue categories. The largest left--right gaps appeared in \textit{Immigration} (18.5\% vs.\ 36.8\%), \textit{Violent Crime} (47.8\% vs.\ 65.8\%), and \textit{Abortion} (17.5\% vs.\ 34.0\%), followed by \textit{Racial \& Gender Inequality} (38.9\% vs.\ 47.4\%) and \textit{Climate Change} (4.9\% vs.\ 12.1\%). Notably, \textit{Violent Crime} also exhibited the highest toxicity prevalence overall across both ideological groups. By contrast, left--right differences were relatively small in \textit{Economy} (9.4\% vs.\ 8.8\%), \textit{Foreign Policy} (27.4\% vs.\ 28.7\%), \textit{Healthcare \& Welfare} (11.7\% vs.\ 13.7\%), and \textit{Democracy} (12.8\% vs.\ 15.3\%).

\begin{figure}[htbp]
    \centering
    \includegraphics[width=1\linewidth]{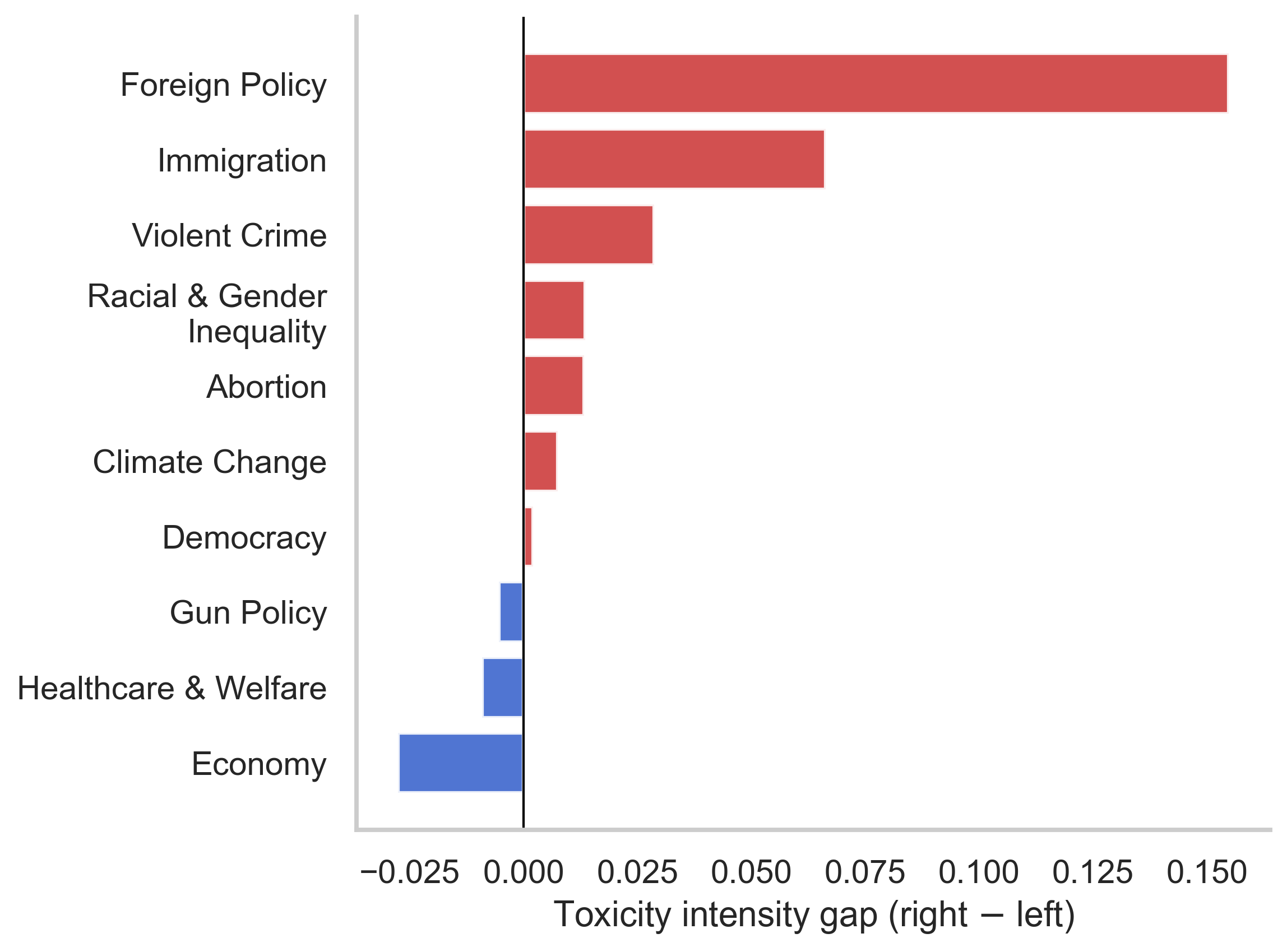}
    \caption{Toxicity Score Gap by Issue and Ideology}
    \label{fig:rq2_issue_ideology_toxicity_score_gap_diverging_bar}
\end{figure}

In terms of toxicity intensity, Figure \ref{fig:rq2_issue_ideology_toxicity_score_gap_diverging_bar} shows the issue-level difference in average toxicity scores between \textit{right}- and \textit{left}-leaning posts. The pattern is uneven and issue-dependent. The largest positive gap appears in \textit{Foreign Policy}, indicating substantially stronger toxicity among \textit{right}-leaning posts on this issue. Positive gaps also emerge for \textit{Immigration}, \textit{Violent Crime},  and \textit{Racial \& Gender Inequality}, and \textit{Abortion}, with smaller positive differences in \textit{Climate Change} and \textit{Democracy}. By contrast, the pattern reverses for \textit{Economy}, \textit{Healthcare \& Welfare} and \textit{Gun Policy}, where \textit{left}-leaning posts show higher toxicity intensity. 

\subsection{RQ3: Psychosocial Dimensions of Toxicity}

To answer RQ3 about understanding the psychological mechanisms driving toxicity, we examined both emotional and moral characteristics of toxic posts.

\subsubsection{Emotion}

We analyzed the emotional characteristics of toxic discourse across campaign issues and ideologies by examining emotional composition within toxic posts. Figure~\ref{fig:rq3-emotion-issue-ideology-radar} shows the distribution of emotion labels across issues and ideological groups (See Appendix~D). 

Across issues, partisan posts were generally more emotional than neutral posts, and toxic discourse was dominated by high-arousal negative emotions. As shown in Figure~\ref{fig:rq3-emotion-issue-ideology-radar}, \textit{Anger} was the most prevalent emotion in most issue domains, often accounting for roughly one-third to over two-fifths of toxic posts. \textit{Fear} and \textit{disgust} also appeared frequently across issues, whereas \textit{joy} and \textit{surprise} remained relatively marginal in most domains.

A key finding is emotional mirroring across ideological groups. Within the same issue, toxic posts from \textit{left}- and \textit{right}-leaning groups often show broadly similar emotional compositions, typically centered on \textit{anger} and followed by \textit{fear}, \textit{disgust}, or the \textit{neutral} label. For example, in \textit{Democracy}, \textit{anger} accounts for 42.4\% of toxic \textit{left}-leaning posts and 41.9\% of toxic \textit{right}-leaning posts, and in \textit{Economy}, the corresponding shares are 42.6\% and 41.3\%. This mirroring suggests that opposing ideological groups often react to the same issue environment with similar emotional repertoires, even when their political positions differ.

At the same time, issue-level differences appear more pronounced than ideological ones. \textit{Foreign Policy} and \textit{Violent Crime} show comparatively stronger \textit{fear}, indicating that toxic discourse in these domains is more strongly shaped by threat and insecurity. By contrast, identity-related issues such as \textit{Racial \& Gender Inequality} and \textit{Abortion} display relatively stronger \textit{disgust}, suggesting a more morally charged emotional profile. Some ideological asymmetries nevertheless remain. For example, \textit{anger} is more prominent in \textit{right}-leaning than \textit{left}-leaning toxic posts in \textit{Foreign Policy} (40.4\% vs.\ 28.9\%) and \textit{Violent Crime} (42.8\% vs.\ 35.6\%).

\subsubsection{Moral Foundation}

We analyzed the moral characteristics of toxic discourse across campaign issues by examining the distribution of moral foundation labels within toxic posts (see Appendix~D).

As shown in Figure~\ref{fig:rq3-moral-issue-ideology-radar}, a key pattern is moral mirroring across ideologies. Within the same issue, toxic \textit{left}- and \textit{right}-leaning posts often emphasize similar dominant moral foundations. For example, in \textit{Immigration}, both sides are structured primarily around \textit{care} (\textit{left}: 32.7\%; \textit{right}: 34.2\%), with \textit{harm} and \textit{cheating} also playing important roles. A similar mirrored structure appears in \textit{Economy}, where both sides are again dominated by \textit{care} (\textit{left}: 31.7\%; \textit{right}: 32.9\%).

Issue-level differences in moral framing also appear more pronounced than ideological ones. For example, \textit{Immigration} is consistently dominated by \textit{care}, reaching 32.7\% of toxic \textit{left}-leaning posts, 30.7\% of toxic \textit{neutral} posts, and 34.2\% of toxic \textit{right}-leaning posts. By contrast, \textit{Violent Crime} is dominated by \textit{harm}, which reaches 44.9\% of toxic \textit{left}-leaning posts, 47.6\% of toxic \textit{neutral} posts, and 49.6\% of toxic \textit{right}-leaning posts. Overall, these patterns suggest that the moral framing of toxic discourse is shaped more strongly by issue context than by ideology alone.

\section{Discussion}

Our findings provide a fine-grained view of discourse during the 2024 U.S. presidential election on X. By combining campaign issue annotation, ideological classification, and psycholinguistic analysis, we move beyond overall toxicity detection to show that online political toxicity was shaped by issue context, ideological alignment, and distinct emotional and moral repertoires. 

\subsection{Harassment Dominants Election Toxicity}

The overall category distribution suggests that toxic election discourse on X was driven primarily by \textit{harassment}, followed by \textit{violence}, whereas \textit{hate} was less prevalent overall. This matters because it indicates that the dominant form of toxicity in our dataset was not explicit hate speech alone, but a broader pattern of antagonistic, interpersonal attack.

One possible interpretation is that this pattern reflects both platform governance and platform interaction dynamics. Prior research shows that online harmful speech persists despite content moderation, and that platforms vary substantially in how they define and govern different forms of harmful speech \cite{gorwa2020algorithmic,singhal2024policies}. At the same time, harassment can function as a socially and politically legible mode of attack that reinforces group boundaries without always taking the form of explicit hate speech \cite{davidson2017automated,marwick2021harassment,rossini_2020_beyond,waseem2016hateful}. In highly polarized environments, such antagonistic expression may be especially compatible with the engagement logic of social media, where conflict, outrage, and attack-oriented discourse can attract attention and spread widely \cite{an_2024_curated,iyengar2019origins,munn2020angry}. 

\subsection{Issue Context Matters}

Our RQ1 findings show that campaign issue context structures toxicity in at least three ways: how often it appears, what form it takes, and how severe it becomes. This extends prior work showing that online hostility and incivility vary systematically across topics and discussion contexts rather than appearing uniformly across political talk \cite{rossini_2020_beyond,salminen_2020_topicdriven,theocharis_2020_the}.

One important takeaway is that toxicity prevalence and toxicity intensity do not always move together, and not all high-toxicity issues are toxic in the same way. Identity-related issues, especially \textit{Racial \& Gender Inequality} and \textit{Immigration}, combine high prevalence with the highest average toxicity intensity. By contrast, several violence-heavy issues, including \textit{Foreign Policy}, \textit{Gun Policy}, and \textit{Violent Crime}, contain substantial amounts of toxic content but comparatively lower average toxicity scores.  One possible interpretation is that different political discussion contexts activate different forms of uncivil and intolerant discourse rather than a single type of toxicity \cite{rossini_2020_beyond,salminen_2020_topicdriven}. There is another possibility that violence-dominated issue domains are more tied to the semantic field of war, crime, weapons, and physical danger, rather than real hostility toward others. However, identity-centered issues are more likely to elicit direct hostile expression toward social groups or out-groups. 

In this sense, aggregate toxicity measures alone may flatten important differences across forms of political hostility. A deeper understanding of toxic discourse, therefore, requires distinguishing between descriptive toxicity, in which violent or harmful language is mentioned semantically without directly targeting others, and more direct forms of hostile expression. Current automated toxicity detection models, including advanced LLM-based models \cite{kivlichan2024upgrading}, still struggle to capture these distinctions at scale. For example, they may not reliably distinguish between language that describes violent events and direct hostile expression. They may also miss causal links that indicate who is attacking whom. Because these models are now widely used in toxicity research, their outputs should be interpreted with caution.

\subsection{Ideological Asymmetries in Toxicity Across Issues}

Our RQ2 findings suggest that ideological asymmetry in toxicity is best understood as a conditional feature of political activation rather than as a fixed property of one side of the spectrum. The contrast between \textit{neutral} and partisan posts indicates that toxicity is more likely to emerge when users are not merely discussing politics, but doing so from a clearly aligned ideological position. In this sense, toxicity appears to be tied less to political discussion itself than to the activation of partisan stakes, identities, and antagonisms in digital space.

At the same time, the asymmetry between \textit{left}- and \textit{right}-leaning discourse is not uniform across issues. Rather than pointing to a stable pattern in which one side is always more toxic, the results suggest that hostility is activated unevenly across issue domains. This implies that ideological toxicity is context-sensitive: different issues appear to mobilize different kinds of grievances, threat perceptions, and political investments, which in turn shape when and how toxicity intensifies.

The category-level results add a further layer of insight. The dominant role of \textit{harassment} suggests that partisan toxicity is most often expressed through direct antagonistic interaction. At the same time, the stronger asymmetry in \textit{violence} indicates that ideological differences are not only about the amount of toxicity, but also about its rhetorical mode. In other words, partisan camps do not simply vary in how toxic they are; they also differ in the kinds of harmful expression they are more likely to mobilize.

Taken together, these patterns suggest that online ideological conflict is better understood as asymmetrically activated hostility under specific issue contexts. What matters is not only which side is more toxic overall, but which issues activate toxicity, for whom, and in what form.

\subsection{Emotional and Moral Mirroring}

Our psycholinguistic findings highlight the emotional and moral structure through which toxicity is articulated, and suggest that toxic political conflict is often rhetorically mirrored: opposing groups may diverge in ideologies while converging in the emotional and moral structures through which they express toxicity. 

Across issues, high-arousal negative emotions consistently play important roles in toxic discourse, and partisan posts tend to be more emotional. More importantly, these emotions are often mirrored between ideological groups. Within the same issue, \textit{left}- and \textit{right}-leaning toxic posts frequently display similar emotional profiles even when they differ in toxicity prevalence or intensity. We use the term \textit{mirroring} here to describe similarity in emotional expression, rather than emotional contagion in a causal sense. The mirrored patterns suggest that opposing ideological groups often respond to the same issue context through parallel emotional logics. In this sense, toxic conflict online may be triggered by mirroring emotions. 

A similar pattern emerges in moral foundations. Within the same issue, ideological groups often rely on the same leading moral framings while attaching those foundations to different political narratives, enemies, or perceived victims. This suggests that partisan content may draw on overlapping moral repertoires while directing them toward competing interpretations of the same issue. However, emotions and moral foundations have been applied differently across issues. This again highlights that the issue's context matters for understanding online toxicity.

\subsection{Limitations}

Several limitations should be considered when interpreting our findings. First, our pipeline cannot fully recover the \textit{target} and \textit{context} of harmful expression. For example, our toxicity measures do not distinguish whether hostility is directed at public figures, private individuals, social groups, or institutions, nor do they capture who is attacking whom. Second, our analysis is based on post-level text rather than interactional relationships. We therefore do not distinguish between in-group and out-group targeting or model how hostility spreads through reply networks and conversational ties \citep{lerman_2024_affective}. Future work could connect discourse-level mirroring with interaction-level polarization. Third, our data likely provide a conservative view of the broader toxicity landscape. Because some highly toxic content may have been removed before collection through platform moderation. In addition, because moderation systems are proprietary, we cannot determine exactly how platform filtering shaped the harmful content that remained visible. Fourth, our findings depend on automated annotation systems, which remain imperfect. These models may still struggle with domain shift, sarcasm, coded language, and politically contextual expression. This concern is especially relevant for emotion and moral foundation analysis. Moral foundation detection remains difficult, and our toxicity analysis also relies on a black-box moderation model. We can observe its outputs, but not the full internal criteria by which toxicity is defined and classified. Future work would benefit from stronger validation and more transparent models.

\section{Ethical Considerations}

This study analyzes publicly available posts from X (formerly Twitter) and focuses on aggregate patterns of toxicity, issue framing, and political ideology during the 2024 U.S.\ presidential election. We do not interact with users or attempt to identify individuals. Because this study examines online toxicity in political discourse, there is a risk of overgeneralizing toxicity as an inherent feature of particular issues or ideological groups. We therefore interpret our findings as descriptive patterns of language observed on one platform during a specific election period, rather than as essential traits of political communities or voters. We caution against using these results for profiling, surveillance, or punitive enforcement targeting specific political groups. Our analysis also relies on automated annotation systems for toxicity, issue, ideology, emotion detection, and moral foundation analysis. These models may encode biases from training data and may misclassify sarcasm, coded language, identity references, or politically contextual speech. We therefore treat model outputs as proxies rather than ground truth, validate key annotations with human coding, and interpret model-based comparisons cautiously.

\section{Acknowledgements}

The authors used generative AI tools to assist with grammar checking, text refinement, and code review. All AI outputs were carefully reviewed and revised by the authors. The authors remain fully responsible for the manuscript. This project was partly supported by NSF (Award \#2331722).

\bibliography{aaai2026}

\clearpage

\clearpage
\onecolumn
\appendix\section{Appendix A. Campaign Issues}

\begin{table*}[htbp]
    \centering
    \resizebox{.86\textwidth}{!}{
        \begin{tabular}{p{0.12\textwidth} p{0.45\textwidth} p{0.43\textwidth}}
            \toprule
            \textbf{\textbf{Issue}} & \textbf{\textbf{Example Keywords}} & \textbf{\textbf{Example Post}} \\
            \midrule
            Economy &
            economy, inflation, unemployment, jobs, wages, GDP, tax, deficit, budget, recession, poverty, middle class, cost of living, small business, trade, manufacturing, minimum wage, investment, IMF &
            I voted down ballot blue including for @crystal\_quade, @LucasKunceMO, and Yes on 3 as well as raising the minimum waging. Thank God I was able to vote early. .@TheDemocrats .@MoDemParty \\
            \midrule
            Healthcare \& Welfare &
            healthcare, health care, insurance, medicare, medicaid, Obamacare, ACA, prescription drugs, hospital, pandemic, covid, mental health, public health, vaccine &
            No one should EVER be forced to choose between paying medical bills or putting food on their table. MAGA Republicans are hell-bent on ripping apart the Affordable Care Act, Medicare and Medicaid! Please SHARE our Healthcare Social Media Kit: https://t.co/WG9YjGNVDS https://t.co/QYyfTkWkwa \\
            \midrule
            Foreign Policy &
            foreign policy, international relations, diplomacy, war, military, defense, troops, NATO, China, Russia, Ukraine, Israel, Gaza, Hamas, Iran, sanctions, allies, terrorism, national security, foreign aid &
            The @StateDept is worried about Israel killing 12 children. What about whenever POTUS Biden bombed that family including children in Afghanistan? That’s the problem the USA needs to mind their own business, period. Israel has the right to protect themselves, they can’t help it \\
            \midrule
            Violent Crime &
            crime, violent crime, homicide, murder, policing, police reform, law enforcement, safety, public safety, criminal justice, justice system, incarceration, prison, defund police, gun violence, fentanyl, drugs &
            Why would you vote for drugs in our state??!? @realDonaldTrump \\
            \midrule
            Immigration &
            immigration, migrant, immigrant, border, asylum, deportation, ICE, wall, DACA, refugee, border patrol, illegal immigration, visa &
            CBS normalizes illegal immigration with 60 Minutes' hour-long infomercial for Kamala Harris https://t.co/ng130yFSnf via @americanwire\_ \\
            \midrule
            Gun Policy &
            gun, firearm, second amendment, NRA, background check, assault weapon, mass shooting, gun control, gun rights, open carry, concealed carry, school shooting, gun safety &
            Nobody wants a Kammunist Kamala Harris gun free world. @KamalaHarris \#KamalaHarris \#KamalaHarris2024 \#KamalaDumpsterFire https://t.co/KmlSWWvZNA \\
            \midrule
            Abortion&
            abortion, reproductive rights, Roe v Wade, pro-choice, pro-life, Planned Parenthood, contraception, reproductive health, women’s rights, bodily autonomy, Dobbs decision &
            Inside Ron DeSantis’s Quest to Trample the Will of Florida Voters on Abortion https://t.co/HF2Ta5nQUr \#YesOn4 \#DeSantisMustGo \#LegalizeIt \#LegalizeAbortion \#ReproductiveFreedom \#ProChoice \#SaveDemocracyVoteBlue \#StopFascism \#StopProject2025 \\
            \midrule
            Racial \& Gender Inequality &
            LGBTQ rights, race, racism, Black Lives Matter, gender equality, trans rights, feminism, discrimination, affirmative action, DEI, inclusion, women’s rights, equity, queer, lesbian, gay, trans, transgender &
            After this election I think the Latino/Hispanic community needs to unpack how they are willing to vote for someone who has no problem disrespecting them behind more conservative laws surrounding abortion, religion, and laws against the LGBTQ+ community. \\
            \midrule
            Climate Change &
            climate change, global warming, environment, carbon, renewable, fossil fuel, clean energy, emissions, green energy, Paris Agreement, sustainability, pollution, wildfire, hurricane, climate policy &
            Five ways a Trump presidency would be disastrous for the climate | US elections 2024 | The Guardian https://t.co/jJcsUkzt4o \\
            \midrule
            Democracy &
            Supreme Court appointments, accusations, misinformation, election legitimacy, election fairness &
            No one wants what Trump and his MAGA Republicans are selling. We want the freedom to make choices for ourselves without government interference. This is America, not North Korea. Just say NO to Trump. Vote YES to Kamala and freedom! \#DemsUnited \#TrumpIsUnfitForOffice https://t.co/0hDFpz5ed5 \\
            \bottomrule
        \end{tabular}
    }
    \caption{Campaign Issue with Example Keywords and Posts}
    \label{tab:campaign_issue}
\end{table*}

\clearpage
\twocolumn
\appendix\section{Appendix B. Descriptive Analysis}

\begin{table}[htbp]
    \centering
    \resizebox{0.5\textwidth}{!}{
        \begin{tabular}{lrrrrrr}
            \toprule
            \textbf{Category} & \textbf{Count} & \textbf{\% of all posts} & \textbf{Min} & \textbf{Max} & \textbf{Mean} & \textbf{SD} \\
            \midrule
            Harassment & 85,787 & 11.81 & 0.37 & 1.00 & 0.69 & 0.18 \\
            Violence   & 32,915 & 4.53  & 0.20 & 1.00 & 0.43 & 0.18 \\
            Hate       & 13,048 & 1.80  & 0.41 & 1.00 & 0.62 & 0.14 \\
            Sexual     & 2,266  & 0.31  & 0.20 & 0.99 & 0.56 & 0.23 \\
            Self-harm  & 544    & 0.07  & 0.28 & 0.99 & 0.51 & 0.18 \\
            Illicit    & 413    & 0.06  & 0.13 & 0.85 & 0.32 & 0.18 \\
            \bottomrule
        \end{tabular}
    }
    \caption{Prevalence and score distribution of toxicity categories}
    \label{tab:descriptive_analysis_toxicity}
\end{table}

\begin{table}[htbp]
    \centering
    \resizebox{0.5\textwidth}{!}{
    \begin{tabular}{lrrr}
        \toprule
        \textbf{Issue} & \textbf{Total posts} & \textbf{Toxic posts} & \textbf{Toxic (\%)} \\
        \midrule
        Democracy                    & 209,497 & 23,685 & 11.3 \\
        Foreign Policy               & 102,068 & 21,466 & 21.0 \\
        Racial \& Gender Inequality  & 20,665  & 8,108  & 39.2 \\
        Immigration                  & 30,919  & 9,042  & 29.2 \\
        Economy                      & 38,767  & 2,171  & 5.6  \\
        Healthcare \& Welfare        & 17,459  & 1,395  & 8.0  \\
        Gun Policy                   & 1,546   & 328    & 21.2 \\
        Abortion                     & 5,319   & 1,286  & 24.2 \\
        Climate Change               & 5,965   & 309    & 5.2  \\
        Violent Crime                & 3,111   & 1,799  & 57.8 \\
        \bottomrule
    \end{tabular}
    }
    \caption{Distribution of Toxic Posts for Each Issue}
    \label{tab:issue_stats_pct}
\end{table}

\rule{0pt}{32\baselineskip}

\pagebreak

\appendix\section{Appendix C. Ideological difference in Toxicity}

\begin{figure}[htbp]
  \centering
  \captionsetup{
    justification=centering,
    singlelinecheck=false,
    width=\linewidth
  }
  \includegraphics[
    width=0.9\linewidth,
    keepaspectratio
  ]{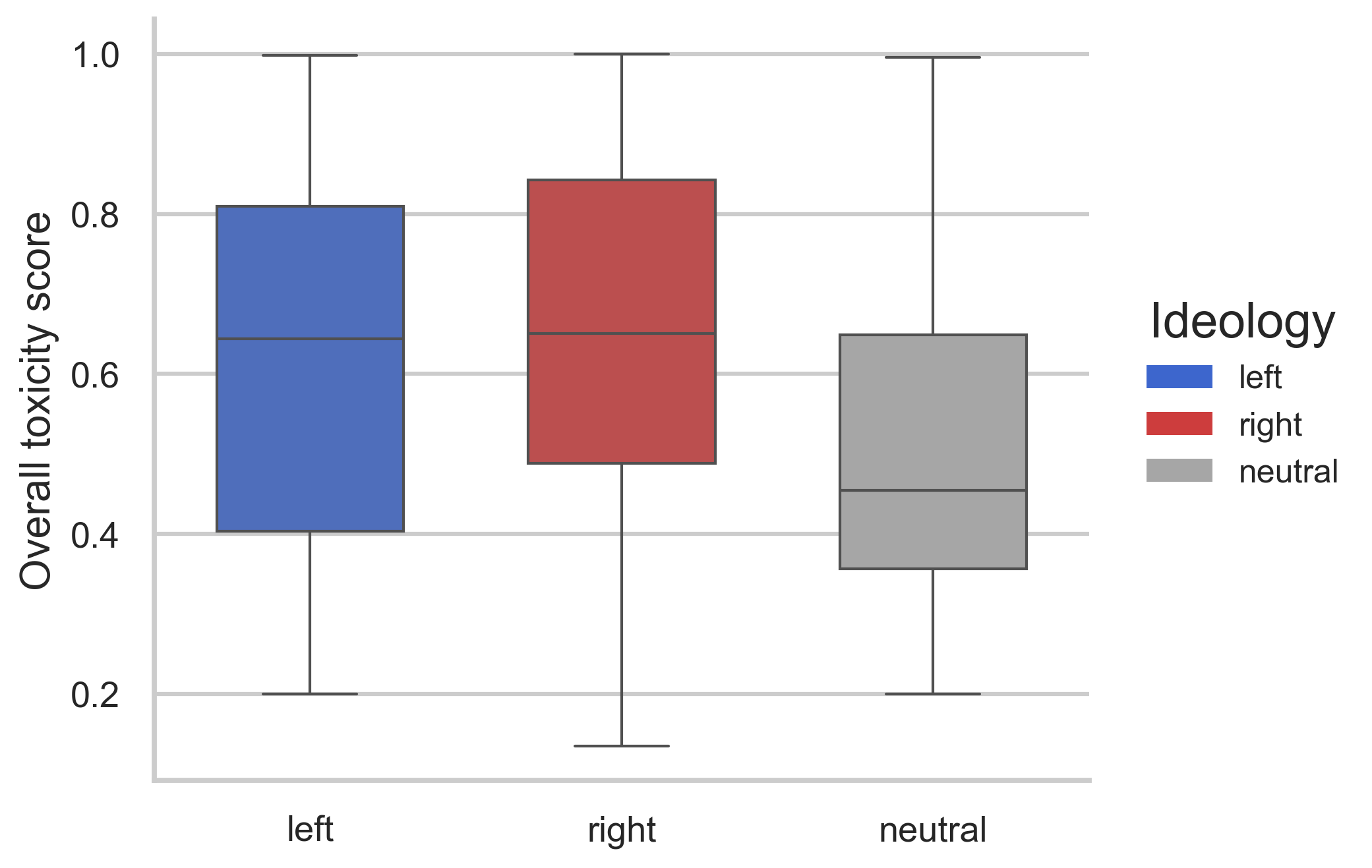}
  \caption{Ideology Toxicity Intensity Boxplot}
  \label{fig:rq2_ideology_toxicity_intensity_boxplot}
\end{figure}

\clearpage
\onecolumn
\appendix\section{Appendix D. Emotion and Moral Foundation Breakdowns by Issue and Ideology}

\begin{figure*}[htbp]
  \centering
  \captionsetup{
    justification=centering,
    singlelinecheck=false,
    width=\linewidth
  }
  \includegraphics[
    width=0.77\linewidth,
    keepaspectratio
  ]{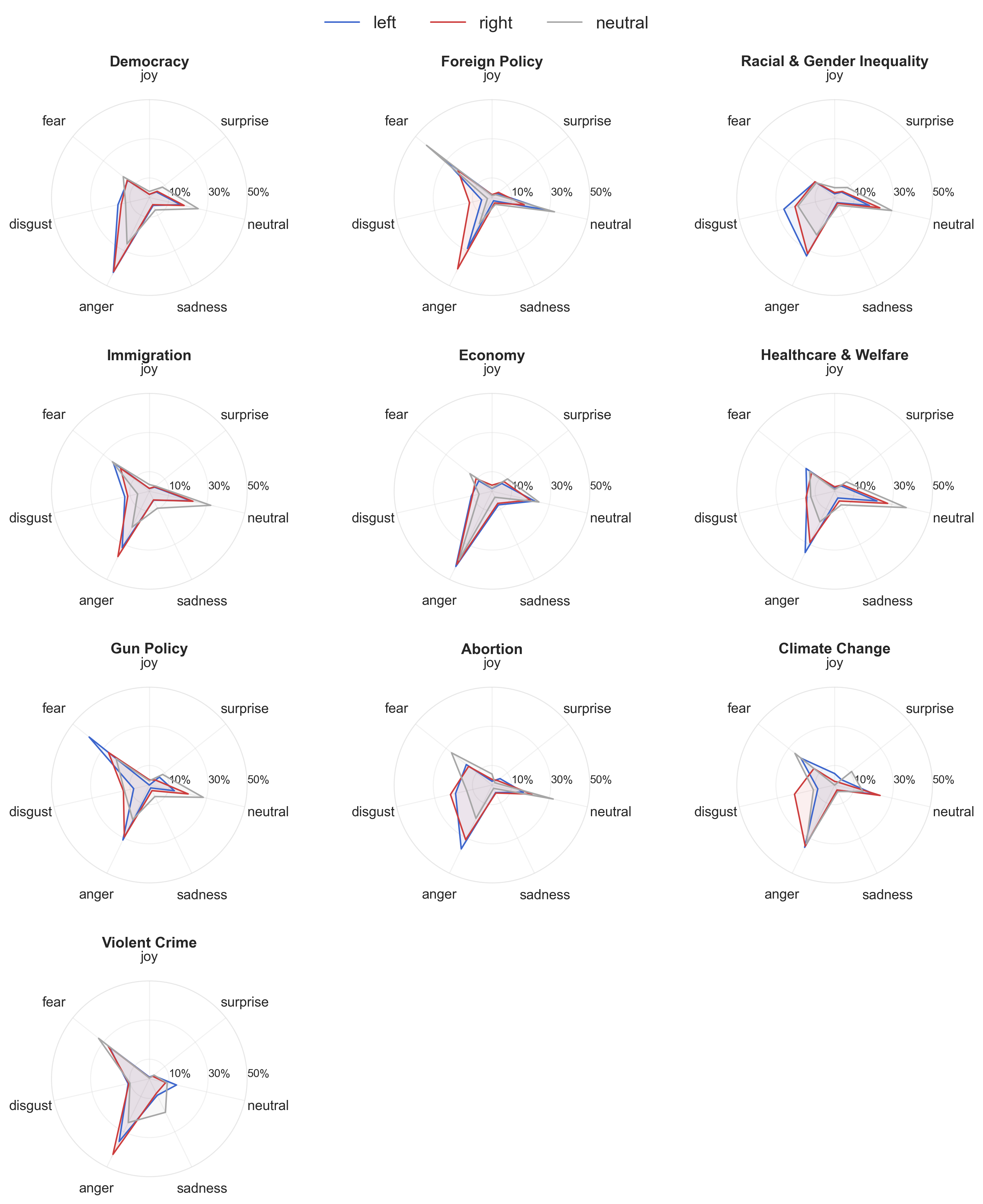}
  \caption{Emotion Label Composition in Toxic Discourse by Issue and Ideology}
  \label{fig:rq3-emotion-issue-ideology-radar}
\end{figure*}

\begin{figure*}[htbp]
  \centering
  \captionsetup{
    justification=centering,
    singlelinecheck=false,
    width=\linewidth
  }
  \includegraphics[
    width=0.77\linewidth,
    keepaspectratio
  ]{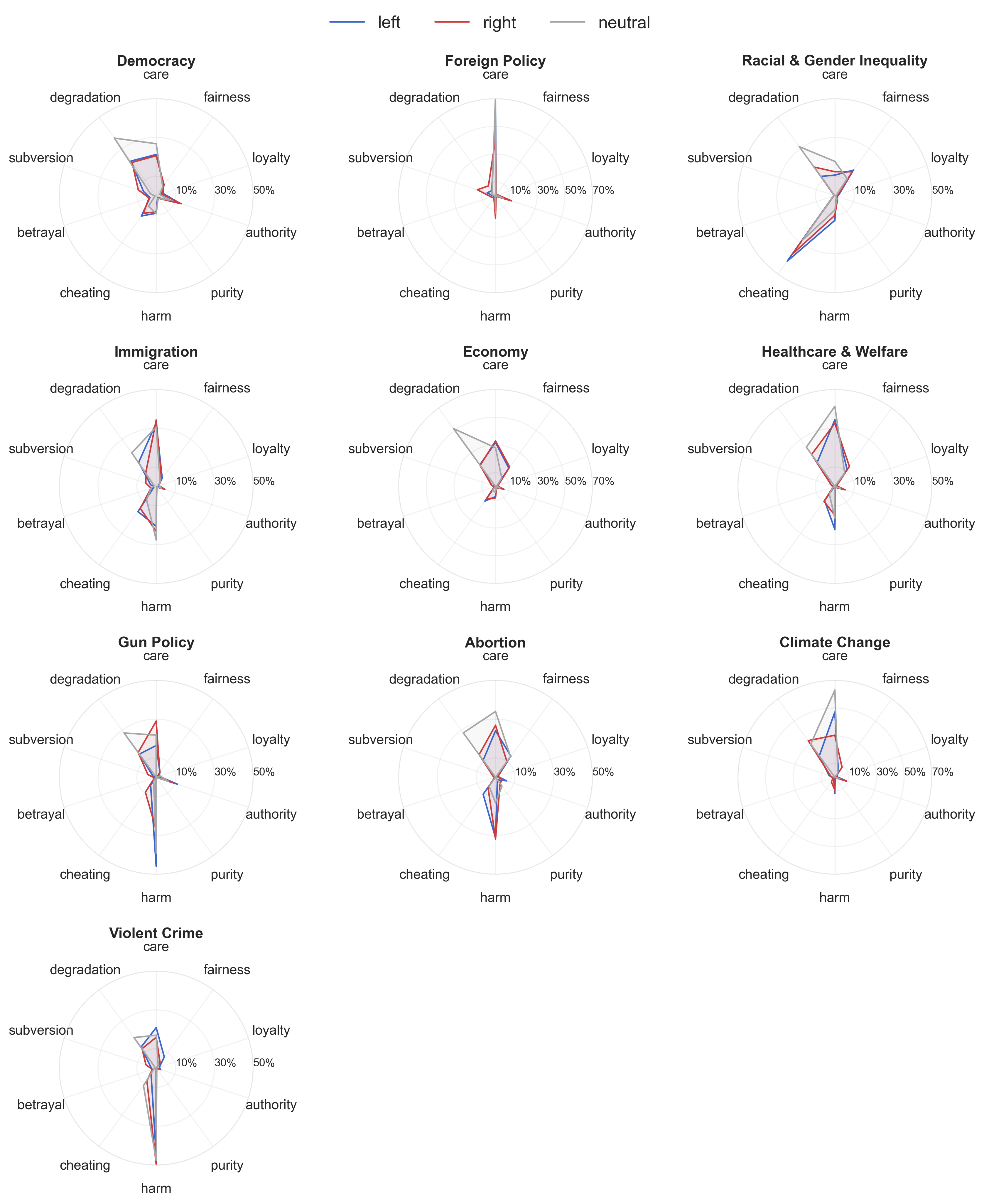}
  \caption{Moral Foundation Composition in Toxic Discourse by Issue and Ideology}
  \label{fig:rq3-moral-issue-ideology-radar}
\end{figure*}

\end{document}